\documentclass[12pt]{article}

\usepackage{epsfig}

\title{Who is the Inflaton?\footnote{Talk given at the Corfu School
    and Workshop of Theoretical Physics 2001}}

\author{Robert Brout\\
\\
{\small Service de Physique Th\'eorique, CP225}\\
{\small Universit\'e Libre de Bruxelles}\\
{\small Bld du Triomphe, 1050 Brussels, Belgium}}

\begin{document}
\maketitle

\begin{abstract}
In the context of the two-fluid model introduced to tame the transplanckian problem of black hole physics, the inflaton field of the chaotic inflation scenario is identified with the fluctuation of the density of modes. Its mass comes about from the exchange of degrees of freedom between the two fluids.
  \end{abstract}

\maketitle

\subsection*{Introduction}

In addition to all else, the major scientific revolutions of the 20th
century, quantum mechanics and relativity, have made it possible to envision
cosmogenesis on a rational basis. Moreover, the theoretical approach that is
envisioned has the potentiality (which has at present to some extent been
realized), of making contact with present day cosmological observations.

\begin{figure}
\begin{center}
\epsfig{file=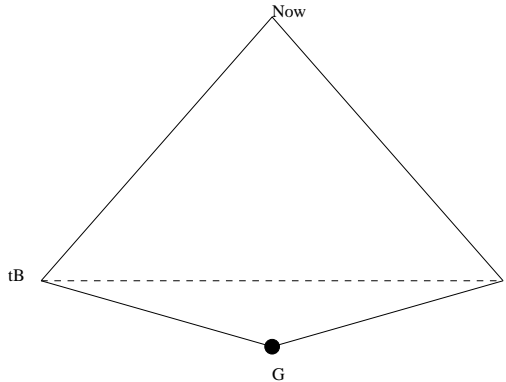,height=5cm}
\caption{\label{fig1} A schematic view of the causality problem}
\end{center}
\end{figure}

The full awareness of this potentiality came in the late 70's and early
80's. It was first pointed out \cite{one} that gravity together with the quantum
theory of matter allowed for the creation of matter {\it ex nihilo}. At the same
time it was shown that creation of matter at a constant rate gave rise to an
exponential expansion of the scale factor \cite{two}, now called inflation and that
this cured the perplexing problems of causality \cite{two} and flatness \cite{three} posed
by the Friedman Robertson Walker adiabatic expansion. A glance at Figure \ref{fig1} is
sufficient to give one a rough appreciation of the causality problem and how
one envisions its solution in terms of creation {\em i.e.} cosmogenesis.
Now one gathers information contained within the backward  light cone as
drawn. At big bang time $\left( t_{B}\right) $, this backward extrapolation
ceases to make sense, if the space-like surface at $t=t_{B}$ contains $1$
degree of freedom per planckian cell.
There must be at least $10^{90}$ such cells since this is the number of
degrees of freedom present within the horizon now, and this number is
conserved during the adiabatic expansion (to ${\cal O}(1)$). Though on the way from $t_{B}$ to now there is some contact among
the degrees of freedom due to scattering, this is far from enough to
account for the observed homogeneity of the CMBR or galaxy distribution (see \cite{four} chapter 8 for details).
Therefore the homogeneity required at $t=t_{B}$ should result from a prior
cause. It is postulated that there is a germ, $G$, from which energy is
created and that the region, which contains this energy, expands
exponentially. This will be realized in the time interval between $G$ and $t_{B}$ if the energy density remains approximately constant within the
region where the energy is created. This is inflation.

\smallskip 

An approach similar to that of \cite{two} was also suggested during this early period
which was based on vacuum energy in the form of the trace anomaly \cite{five}.
Whereas these early attempts were based essentially on production of kinetic
energy, be it in the form of particles \cite{two}, black holes \cite{three} or zero point
energy \cite{five}, a new tactic was introduced in the early 80's which relied on
potential energy. The first idea in this vein was latent heat released in a
first order phase transition \cite{six}.
It was postulated that the universe cooled down from a prior state of high
temperature whereupon at some point the matter contained within would
condense. But it was recognized at once \cite{six} that this approach could not
work. The condensed phase would be delivered in bubbles which would be swept
out by the expansion too fast to allow for their coalescence. In addition I
find that such a scenario would be epistemologically dissatisfying. If one
is out to solve the causality problem, then one would hope that the cause in
question is the cause of the universe i.e. one would like cosmogenesis and
inflation to go together. To push the ultimate problem further 
backwards in
time is, in a sense, to escape the problem. 
But this is merely my subjective
opinion.

Fortunately, this approach could be modified through the invention of ``slow
roll'' mechanisms \cite{eight,nine}. This removed the problem of non coalescence of
bubbles since there would be time enough. Moreover in Linde's hands \cite{ten}
chaotic inflation came into being. We shall not here enter into details of
how Linde envisages fluctuations within fluctuations, but simply give that
part of the idea which we shall adopt in the foregoing. (We also refer the
reader to \cite{three} for further consideration of this question of fractalizing
the fluctuation). The point of view we shall take here is that there is a
state of rest in which there is no universe. But because of quantum
mechanics the ``stuff'' in this space fluctuates. The stuff is complicated
in its nature. Only at the long distance scale (and here we adopt the
traditional point of view that the only fundamental scale is planckian) does
usual field theory apply. The short distance scale is described by 
a ``planckian soup'',
the present great unknown of physics. It may involve higher dimensions
and/or non-commutative geometry and/or strings and/or black hole
fluctuations and/or... 
Whatever, we shall only suppose that it is described
by a density or, if one wishes, a scale which is planckian. The
abovementioned state of rest is interpreted variationally. The mean density
in this state is postulated to be the state of minimum energy density in
which there is no expansion of the scale factor in the metric that describes
this state. We shall postulate that this space time is flat {\em i.e.} $\langle
T_{\mu \nu }\rangle =0$, $T_{\mu \nu }$ being the energy momentum tensor of
everything.\footnote{Fran\c{c}ois Englert has suggested to me that I could well let
up on my stringent assumptions and introduce a small fundamental
cosmological constant to describe this state. This would not do injustice to
the sequel. For the novice I conform to what is most simple and most natural
(perhaps ?).}

\smallskip

This quiescent state of things is conceived as metastable. Fluctuations that
are small both in amplitude and extension will regress and leave no
permanent effects. But now comes this wonderous conjunction of quantum
mechanics and relativity.

In the hamiltonian formulation of general relativity, the total energy
density of gravity + matter sums to zero.

Instrumental in the realization of this fact is that the kinetic energy
carried by the motion of the scale factor, $a$, is negative ({\em i.e.} in the
hamiltonian density there is a term in $\left( -H^{2}\right) $ where $H$ is
the Hubble constant: $H=\left( \dot{a}/a\right) $, where dot is derivative
with respect to proper time). Thus the positive energy density of a
fluctuation of matter density is necessarily accompanied by $H\neq 0$, so as
to keep the energy density equal to zero. This constraint is strict; even
in the future when fundamental theory will relegate the Einsteinian theory
of gravity to successful long scale phenomenology it will probably survive.
It is due to the invariance of the total action against reparametrezation of
the temporal coordinate and one would be loath to give up such a fundamental
principle in whatever guise it will express itself in future theory.

It is herein that opens the possibility of permanent effects of a
fluctuation. If this latter is large enough in amplitude (to allow for a
systematic temporal development) and in extension (so as to render spatial
gradients sufficiently small in energy as compared to the total energy of
the fluctuation) then it will be seized upon by the scale factor and will
expand exponentially immediately after its formation. After a while the
fluctuation will nevertheless regress and one postulates its energy is
converted to ordinary quanta whereupon the entropy so produced remains
constant and FRW adiabatic expansion ensues.

At the present time, the formal development of the above scenario is
executed in terms of a scalar field. All the aforementioned concepts fall
neatly into this development and furthermore, detailed quantitative analysis
has shown that it is possible to account for presently observed density
fluctuations both of the anisotropy of the CMBR and the distribution of
galaxies. For details see \cite{four}, chapter 8, and the lectures of Lazarides at this summer school and workshop at Corfu 2001.

It is my purpose here to dwell on the possible nature of this fundamental
fluctuation, often called the ''inflaton'' field albeit in highly
speculative fashion. Who is he, this inflaton ?

My inspiration has been in great measure drawn from the experience that has
been garnered from the theory of black hole physics over the past decades
and which finds it expression, in the 2-fluid phenomenology that has been
developed in recent years \cite{eleven}.

\subsection*{The Cosmological Fluid {\em vs} its Material Counterpart}

The theory of Hawking evaporation \cite{twelve} from black holes, (and
we take as the simplest example s-wave emission form Schwarzchild black
holes) encounters the transplanckian problem. In the original version of the
theory, Hawking studied free field propagation in the background Schwarzchild
metric. He showed that a vacuum fluctuation in the vicinity of the horizon,
as it propagates towards spatial infinity, is possessed of an amplitude
which in part becomes that of a physical on massshell quantum due to the
changing metric it experiences in its voyage. The distribution of these
quanta is thermal, at least for frequencies of the order of or larger than
the inverse horizon distance $\sim m_{pl}^2 / M$, where $M$ is
the mass of the black hole. In the derivation of this remarkable phenomenon
one appeals to configurations of fields which are localized to within
fantastically small distances of the horizon $\Delta r\sim Me^{-\alpha M}$
 (where $\alpha ={\cal O}(1)$).  Here and from now on we work in planckian units $m_{pl}=1$. Correspondingly the energies involved in such configurations are $%
\sim M^{-1} e^{\alpha M}$ and this, for a macroscopic black
hole where $M \sim 10^{60}$, is ridiculously large.

Since gravitational interaction grows with the product of the energies of
the interacting entities it is seen that the modes which contribute to
Hawking radiation interact extremely strongly in the regions from whence
they come in order to deliver their asymptotic flux. Therefore the picture
of free mode propagation breaks down. Unruh \cite{thirteen} has shown nevertheless that
this does not kill the effect. All that one requires is a condition that had
been previously proposed by Jacobson \cite{fourteen}, to wit : at some finite distance
from the horizon one still has vacuum conditions. A likely distance is $%
O\left( 1\right) $ {\em i.e.} $m_{pl}^{-1}$ . Unruh showed that if one tinkered
with the dispersion relation, $\omega \left( k\right) $, in a manner similar
to that of phonons in fluids where $\omega $ cuts off at $k=\left(\mbox{\rm
interparticle distance}\right) ^{-1}$, then Jacobson's condition is met {\em i.e.}
Hawking radiation is robust. Unruh's analysis was numerical, but it has also
been demonstrated analytically \cite{fifteen,sixteen}.

Since then, other models have been proposed, one based on a
deformed commutator \cite{seventeen} and another which takes into account scattering
between incoming and outgoing modes \cite{eighteen}.
 In these latter cases backward
extrapolation of the emitted Hawking quantum results in its disappearance
once it enters into a region which is localized around the horizon, which
one supposes is given on the planckian scale (at least in \cite{seventeen}; in \cite{eighteen}
this specification is more complicated). In both cases Hawking steady state
radiation in asymptotia is maintained whereupon the picture emerges that the
quanta boil off from a horizon region, {\em i.e.} there is some kind of planckian
soup which serves as a reservoir of ``cisplanckian'' quanta.

All of this to motivate a two-fluid picture of space-time and the
 matter in
it, wherein the large length scale physics is described by conventional
field theory, modes (and their interactions) which propagate in a fluid
which describes the physics at small length scale. The scale at which the
dichotomy takes place is the cut-off of modes, hence proportional to $n^{-%
\frac{1}{3}}$ where $n$ is a parameter which describes the density of modes.
We expect $\langle n\rangle ={\cal O}\left( m_{pl}^{3}\right) $.

\smallskip Our experience in black hole physics has emboldened us to take
this picture seriously. Hawking radiation is, in no small measure, as solid
as thermodynamics (for the eternal black hole this statement is based on the
periodic structure of Green's functions in imaginary time in Schwarzchild coordinates in the region of space-time where this description is
appropriate. It is also a necessary concomitant to the elimination of
unphysical singularities in $T_{\mu \nu }$ in the region of the horizon). We
know that the mode picture must fail once we reach planckian scales owing to
mode-mode interactions so that beyond that scale, strong interaction
physics take over. And this we do not how to formulate. But the one acid
test where quantum field theory is confronted with strong effects of
gravity, the black hole, has taught us that we can account correctly for a
phenomenom which is solidly established, Hawking emission, using this
cut-off low momentum sector whilst remaining non committal concerning the
nature of the planckian soup. And further from \cite{seventeen} and \cite{eighteen}, this soup
serves as reservoir for the low momentum modes.

What does this have to do about the inflaton and the primordial cosmogenetic
germ? The answer lies in quantum mechanics. The collective variable, $n$,
appears in our description of the wave function of everything. Therefore it
fluctuates. Moreover it fluctuates locally. It is a field $n=n\left(
x,t\right) $. Our proposition is that at least one piece of the physics that
is described phenomenologically by an inflaton field is the fluctuation of $n$. There may be more to the inflaton than $n$ alone, 
due to the dynamics of
matter. For example supersymmetry and its breaking may have important
physical repercussions on how $T_{\mu \nu }$ fluctuates. But, given our
present appreciation of gravity in its quantum context, fluctuations of $n$
are inevitable. Whether this is ultimately determinant, (i.e. whether the
fluctuation of $n$ is the germ) is a quantitative question which requires
further research. In the foregoing paragraphs I shall simply indicate that
the fluctuation of $n$ has the qualitative attributes that one attributes to
the inflaton field. Hopefully more quantitative support (or negation) will
emerge in subsequent work.

One very important question is that of the cosmological constant $\Lambda $.
I am assuming $\Lambda =0$ in the initial quiescent universeless state (at
least in the region of space-time that holds our universe). I am therefore
waiving the interesting problem posed by the existence of $\Lambda $ now,
which appears to the same order of magnitude of the present energy density $\sim 10^{-122}m_{pl}^{4}$. This seems more like a tuning
problem than the fundamental issue we are facing here where by all rights
one might have expected $\Lambda \sim m_{pl}^4$ .

The answer to why $\Lambda =0$ at the beginning may be presumed to be hidden
in the variational principle which fixes $\langle n\rangle $ {\em i.e.} in the
quiescent state one has $\partial \langle \epsilon \rangle /\partial n\left|
_{n=\langle n\rangle }\right. =0$ where $\langle \epsilon \rangle $ the mean
energy density. One must find out why $\langle \epsilon \left( n=\langle
n\rangle \right) \rangle =0$.

The question is nigh to impossible to answer. This is because the energy of
space-time in this 2 fluid picture takes the form
\begin{equation}
\epsilon \left( n\right) =\epsilon _{{mod}e}\left( n\right) +\epsilon
_{soup}\left( n\right) + \mbox{\rm (interactions between soup and modes)}
\label{Eq1}
\end{equation}
We only know something about $\epsilon _{{mod}e}\left( n\right) $ {\em e.g.}
the zero point contribution is $\sim n^{\frac{4}{3}}$. Of all the rest we
are ignorant.

There are two points of view which one can adopt in setting $\Lambda =0$.
One is the time honored procedure of blind substraction. This can be couched
in somewhat more elegant terms as a sort of Aristotelian principle of
absolute rest, to wit: gravity does not respond to the configuration of
fields which is at the rock bottom of $\epsilon \left( n\right) $. It only
responds to the fluctuations {\em i.e.} to $\tilde{\epsilon}\left( n\right)
=\epsilon \left( n\right) -\epsilon \left( \langle n\rangle \right) $ where $%
\epsilon \left( \langle n\rangle \right) $ is the value of $\epsilon \left(
n\right) $ at its minimum.

The alternative possibility is that $\Lambda =0$ is a principle, say as
fundamental as the principle of equivalence. This would fix $\epsilon \left(
\langle n\rangle \right) =0$. The implication is that there is a balancing
out of parameters which appear in the various terms of (\ref{Eq1}), which
arranges things just so. (Using foam models it is easy to make such
constructions, but they are on no more nor less fundamental grounds that the
more usual {\em ad hoc} subtraction).

Be that as it may, an expansion about $\langle n\rangle $ leads to
\begin{equation}
\tilde{\epsilon}\left( n\right)  \approx \frac{1}{2}B\left( \delta n\right) ^{2}
\label{Eq2}
\end{equation}
where $\delta n=n-\langle n\rangle $ ; $B=\partial ^{2}\epsilon /\partial
n^{2}\left| _{n=\langle n\rangle }\right. $

An important point is to realize the difference in consequences of (\ref{Eq2})
when one compares the cosmological fluid to a stable material fluid where 
(\ref{Eq2}) would also apply (say to a superfluid at zero temperature). The
difference lies in the conservation of the elements (atoms) which make up the
latter and their non conservation in the former. (For example, in the naive model of foam made up of instantons, in virtue of the very word, these
entities come and go).

At this point, it is meet to set out an important caveat. It has been argued
in \cite{eleven} that $\Lambda =0$ in virtue of stability. This argument is
insufficient because it is based on conservation of the number of atoms. It
is worth setting out the argument since it then allows one to appreciate why
density fluctuations are massless in the case of material liquids, but massy
in the cosmological case.

Consider a superfluid at zero temperature. It is stable {\em i.e.}
\begin{equation}
-p=\left( \partial E/\partial V\right) _{N,T=0}=0  \label{Eq3}
\end{equation}
An observer inside the fluid uses the grand ensemble to describe his physics {\em i.e.} within the volume, $V$, which is within his scope of observation,
particles come and go and this is accounted for by making use of the energy
in $V$ which we will denote by $\tilde{E}\left( V\right)$
\begin{equation}
\tilde{E}\left( V\right) =E\left( V\right) -\mu \langle N\rangle 
\label{Eq4}
\end{equation}
where $\langle N\rangle $ is the average value of $N$ in $V$ and $\mu $ =
chemical potential ; $\mu =\left( \partial G/\partial N\right) _{p,T}$ and
at $T=0$, $G=E+pV$. Also since $p=0$ in virtue of stability we have $G=E$,
whence 
\begin{equation}
\tilde{\epsilon}=\tilde{E}/V=\epsilon -\mu n=0  \label{Eq5}
\end{equation}
The conditions $\tilde{\epsilon}=p=0$ are translated into zero cosmological
constant.

The argument is not applicable to the cosmological fluid since $N$ is not
conserved whence $\partial /\partial V\left|_{N}\right. $ is not a physical
variation. Rather $p=-d E/d V$ at $T=0$ {\em i.e.} as $V$ varies, $N$ can vary with
it to keep the density $n$ (whose average is $\langle N\rangle /V$) at its
optimal value. In other words, as $V$ varies particles can pop in and out of
the vacuum so as to keep $\langle N\rangle /V$ fixed. Therefore stability
does not imply $p=0$.

The true variational principle in this case is
\begin{equation}
\mu =\left( \partial E/\partial N\right)_{V}=0  \label{Eq6}
\end{equation}
One is generally led to believe that $\mu =0$ when number is not conserved
is a negative statement. If you don't need it then it is zero. In the
present case, this is also true, but it is a more powerful statement than
usual in that (\ref{Eq6}) fixes the density to be optimal. It also implies $%
\partial \epsilon /\partial n\left| _{n=\langle n\rangle }\right. =0$. And
this is a far cry from $\epsilon \left( n=\langle n\rangle \right) =0$.

What (\ref{Eq6}) does imply is that the mean energy of the ground state of the
cosmological fluid is delivered in the form of a cosmological constant.
Indeed
\begin{equation}
\mu =\langle \epsilon \rangle +p=0  \label{Eq7}
\end{equation}
means $p=-\langle \epsilon \rangle $. In general relativity this means $%
\langle T_{\mu \nu }\rangle =\Lambda g_{\mu \nu }$ where in this case $%
g_{\mu \nu }$ is flat. But, to repeat, one must further to prove $\Lambda
=0$, and this is a great problem which continues to be vexing.

Let us now continue with (\ref{Eq2}) and show that it invites the proposition
that $\delta n$ is a massy scalar field which describes density fluctuations
in the cosmological fluid, hence having the property that one associates
with the inflaton. The physics, once more, is brought out by comparison with
the material fluid.

Because $N$ is conserved in the material fluid density fluctuations have no
zero Fourier component. Thus there must be no physical manifestation of the $%
k=0$ component of $\delta n\left( r,t\right) $, whence $\omega \left(
k=0\right) =0$. If this is the terminal point of a continuous spectrum than
the mass of the field $\delta n\left( r,t\right) $ is zero. The way this is
realized is to compute the pressure gradient that is induced by a density
gradient $\nabla p=B\nabla n$. This pressure gradient causes acceleration of
the fluid element ({\em i.e.} temporal variations in the fluid current $j$). Using
the equation of continuity, $\partial n/\partial t+\nabla\cdot j=0$,
then straightforwardly one produces the wave equation for sound with $\omega
^{2}\left( k\right) \sim Bk^{2}$.

What is important is that all the energy of a fluctuation is in the spatial
gradients plus the acceleration that these induce in the form of kinetic
energy {\em i.e.} $ E \sim (\nabla n)^{2} + \frac{1}{c^{2}} \dot{n}
^{2} $ for small fluctuations where $c$ is the
velocity of sound, $c \sim \sqrt{B/m}$.

\begin{figure}
\begin{center}
\epsfig{file=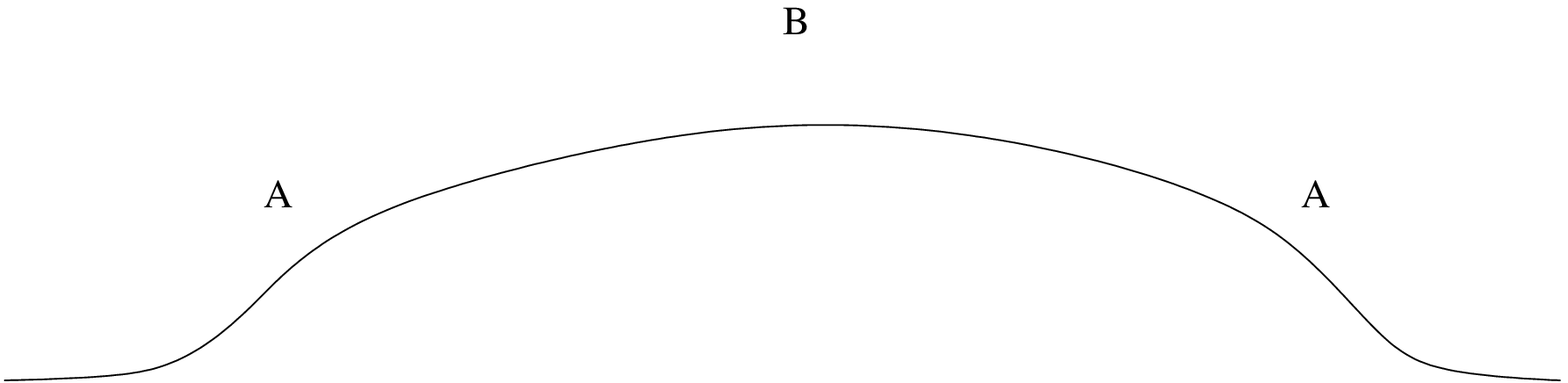,height=2.5cm}
\caption{\label{fig2}}
\end{center}
\end{figure}

Now consider the cosmological fluid. Here the $k=0$ component does exist.
Thus one can have a fluctuation profile which looks like Figure \ref{fig2}.
This is the form one is talking about in inflation and indeed it is how one
thinks about the germ. And there is no reason in the world why such
fluctuations of $n$ should not exist. In addition to the usual energy lodged
in gradient terms as in regions $A$, there is present a flat region $B$
where the gradient vanishes or is small. Nevertheless its energy exists. It
is pulled out of the quiescent configuration because absolute quiescence is
not a quantum concept; $\left[ \epsilon \left( n\right) -\epsilon \left(
\langle n\rangle \right) \right] \neq 0$ can come about by changing the
density of modes and or soup in a region without recourse to $\nabla n\neq 0$
in that region. Thus the term $\frac{1}{2}B\left( \delta n\right) ^{2}$ can
and does contribute to $\frac{1}{2}m^{2}\varphi ^{2}$ where $\varphi $ and $m
$ are a scalar field (the inflaton) and its mass.

Of course gradients will exist in any finite size fluctuation and they will
carry energy. On grounds of covariance their action will be proportional to $
g^{\mu \nu }\partial _{\mu }\varphi \partial _{\nu }\varphi $. To calculate
the relative weights of these kinetic terms as compared to the potential
term is a dynamical problem which must be addressed. 

To assert that the inflaton is $\delta n$ is premature. Various models must
be tested to see if the parameters one comes by are acceptable. But it is
fairly clear that whatever the inflaton turns out to be, the fluctuation of
the fundamental ``stuff'' which we now contemplate as the point of departure
for the formulation of quantum gravity must play some r\^{o}le. It may be
one element of a highly complex process involving any one of the number of
candidates presently considered at the foundation of matter and gravity, or
for all we know it may be the whole story.

We close this rather unsatisfactory speculative discourse with some
additional remarks.

1) Since $n$ is proportional to the density of modes (essentially $n^{\frac{1%
}{3}}$ is the cut-off), fluctuations of $n$ will cause not only fluctuations
of mode density. Some of these fluctuations will end up as those which one
calls upon to account for present fluctuations of the CMBR and structure.
Another part will cause the production of quanta due to the nonadiabatic
variation $\left( \delta \dot{n}\neq 0\right) $. Since $\delta n$ regresses,
one obtains particle production. This is in analogy to production of phonons
induced by the finite rate of change of density in a solid.  Once more
details are required to see if this is the source of the entropy we observe.

2) One must not lose sight of alternatives to the inflaton scenario. Hawking
has investigated black hole pair fluctuations in vacuum. It is not
impossible that there are rare abundant coherent fluctuations which will be
seized upon by the scale factor and which will inflate just as the
fluctuation of $\varphi $ inflates. This is but one example. To be sure the
scenario of chaotic inflation has to some extent ``won its spurs''. But
fundamental cosmology is essentially speculative in character. Any argument
that works is an argument of sufficiency, not of necessity. So I conclude,
it is too easy to be complacent. We are probably going to meet no end of
surprises before we can reach some consensus of opinion on just what it is
that is responsible for the birth of our universe.

\section{Added note}
After the completion of this text, I analyzed somewhat further the potential energy of the fluctuations of the mode density by separating it into two parts:
\begin{description}
\item{1.} the ``acoustic'' contribution generated by pressure gradients wherein total number is conserved,
\item{2.} a source term  occasioned by exchange of a degree of freedom between the ``planckian soup'' and the ``vapor of modes''.
\end{description}
When the soup is modeled by a collection of black holes (or, in general, entities with horizons), then the source generates an inflaton mass of order of magnitude
$$
\mu^2 \approx \exp( - \vert \Delta S\vert) m_{Pl}^2
$$
where $\Delta S$ is the entropy change of the black hole that is induced by the exchange. For Schwarzschild black holes this is given by $
\Delta S = 8 \pi M \bar \omega$, $M=$ mass of the black hole, 
$\bar \omega=$ mean energy of modes. One expects both $M$ and $\bar\omega$ to be ${\cal O}(m_{pl})$ whereupon one finds $\mu = {\cal O}(10^{-5}) m_{pl})$. This is the order of magnitude required by the analysis of present day cosmic fluctuations. N.B.: The above estimate is very crude and can easily change by several orders of magnitude due to its sensitivity on $M$.

\end{document}